\pdfoutput=1
\documentclass[aps,10pt,twocolumn,superscriptaddress]{revtex4-1}
\usepackage{graphicx}
\usepackage{amsmath,amssymb}
\usepackage{txfonts}
\usepackage{bm}
\usepackage[usenames,dvipsnames]{xcolor}
\usepackage{multirow}

\makeatletter
\newcommand*{\balancecolsandclearpage}{%
  \close@column@grid
  \clearpage
  \twocolumngrid
}
\makeatother

\begin{document}

\title{Theory of spin loss at metallic interfaces}
\author{K. D. Belashchenko}
\affiliation{Department of Physics and Astronomy and Nebraska Center for Materials and Nanoscience, University of Nebraska-Lincoln, Lincoln, Nebraska 68588, USA}

\author{Alexey A. Kovalev}
\affiliation{Department of Physics and Astronomy and Nebraska Center for Materials and Nanoscience, University of Nebraska-Lincoln, Lincoln, Nebraska 68588, USA}

\author{M. van Schilfgaarde}
\affiliation{Department of Physics, Kings College London, Strand, London WC2R 2LS, United Kingdom}

\date{\today}

\begin{abstract}
Interfacial spin-flip scattering plays an important role in magnetoelectronic devices. Spin loss at metallic interfaces is usually quantified by matching the magnetoresistance data for multilayers to the Valet-Fert model, while treating each interface as a fictitious bulk layer whose thickness is $\delta$ times the spin-diffusion length. By employing the properly generalized circuit theory and the scattering matrix approaches, we derive the relation of the parameter $\delta$ to the spin-flip transmission and reflection probabilities at an individual interface. It is found that $\delta$ is proportional to the square root of the probability of spin-flip scattering. We calculate the spin-flip transmission probability for flat and rough Cu/Pd interfaces using the Landauer-B\"uttiker method based on the first-principles electronic structure and find $\delta$ in reasonable agreement with experiment.
\end{abstract}

\maketitle

Spin transport at metallic interfaces is an essential ingredient of various spintronic device concepts, such as giant magnetoresistance (GMR) \cite{GMR,Bass,Bass2015}, spin injection and accumulation \cite{Johnson}, spin-transfer torque \cite{Ralph}, and spin pumping \cite{pumping}. Spin-orbit coupling (SOC) enables some device concepts, such as spin-orbit torques in ferromagnet/heavy-metal bilayers \cite{Miron,Liu} and spin current detection based on the inverse spin-Hall effect \cite{ISHE} in spin-caloritronic devices \cite{caloritronics}. Interfacial spin-orbit scattering affects spin transport in GMR multilayers \cite{Bass,Bass2015}, spin pumping \cite{Jaffres,Chen2015}, spin injection \cite{Rashba}, and Gilbert damping \cite{Kelly}. It contributes to the spin relaxation in metallic films \cite{Long1,Long2,Long3} and to the magnetoanisotropies in the resistance of magnetic miltilayers \cite{Kobs}, tunnelling conductance \cite{Gould,Chantis,Moser,Park}, and Andreev reflection \cite{Zutic-AR,Hogl}, which are especially large when the magnetic electrodes are half-metallic \cite{Burton,Hogl}. Interfacial spin-flip scattering can also appear due to spin fluctuations \cite{Zhang}.

In the absence of interfacial spin-flip scattering, spin transport in magnetoelectronic circuits can usually be described using the circuit theory \cite{Brataas.Nazarov.ea:PRL2000,Brataas.Nazarov.ea:EPJB2001,Brataas.PhysRep}. In the presence of SOC, the spin current is not conserved at the interfaces. Absent a complete theory, interfacial spin-flip scattering has been described by introducing a fictitious bulk layer of thickness $t_I$, resistivity $\rho_I$, and spin-diffusion length $l^I_{sf}$, and using the parameter $\delta=t_I/l^I_{sf}$ to characterize ``spin memory loss'' at the interface \cite{Bass,Bass2015,Baxter,Manchon,Kelly}. The parameter $\delta$ was measured \cite{Bass,Bass2015} for multiple interfaces by mapping the experimental current-perpendicular-to-the-plane magnetoresistance data, for spin valves with multilayer insertions, to the phenomenological Valet-Fert model \cite{VF1993}. However, the relation of the parameter $\delta$ to the scattering properties of an individual interface is not known. Moreover, this description of an interface is generally incomplete, because the spin-flip transmittance and the reflectances on two sides are all independent parameters. For example, the spin-flip reflectance is relevant for spin injection \cite{BGW} and for the interface-induced spin relaxation in a spin reservoir \cite{Long1,Long2,Long3}. The existing formulations \cite{Fert-Lee,Rashba,Barnas} including only one interfacial spin-relaxation parameter are, therefore, also incomplete.

In this Letter we apply the scattering matrix and the generalized circuit theory approaches to establish the correspondence between the phenomenological parameter $\delta$ for a nonmagnetic interface, as extracted from GMR-like measurements, and the calculable spin-resolved transmittance and reflectance properties of an individual interface. The latter are calculated from first principles for the Cu/Pd interface. The theory provides a complete framework for including interfacial spin-flip scattering in magnetoelectronic devices.

\paragraph{Valet-Fert theory.}

The layer thicknesses in the typical measurements \cite{Bass,Bass2015} are about 3 nm; the resistance of each individual layer is at least a few times smaller than the resistance of each interface, as long as nominally pure materials are used. For example, the area-resistance products of a 3-nm layer of nominally pure Pd and of the Cu/Pd interface are about 0.14 and 0.45 f$\Omega\cdot$m$^2$, respectively \cite{Bass}. Therefore, in the following we treat the problem under the assumption that the bulk resistances are negligibly small compared to the interface resistances. This simplifies the expressions and does not affect the result to first order in spin-flip scattering rates \cite{supplement}.

To facilitate comparison with scattering theory, it is convenient to consider a periodic multilayer in which the FN$_1$(N$_2$N$_1$)$_\mathcal{N}$ block repeats itself. Here F is a ferromagnetic layer, N$_1$ and N$_2$ are two different non-magnetic layers, and we are interested in the properties of the N$_1$/N$_2$ interface. Describing an interface as a bulk interlayer, we solve the Valet-Fert equations \cite{VF1993} in the multilayer for parallel and alternating antiparallel configurations using the transfer-matrix approach. Taking the limit in which the resistance is dominated by and spin-flip scattering is present only at N$_1$/N$_2$ interfaces, we find a simple expression for the magnetoresistance:
\begin{equation}
\Delta R = R_{AP}-R_{P} = \frac{(\beta r^*_F)^2}{r_I} \frac{\delta}{\sinh m\delta},\label{drVF}
\end{equation}
where $m=2\mathcal{N}$ is the number of interfaces, $\beta=(\rho_\downarrow-\rho_\uparrow)/(\rho_\uparrow+\rho_\downarrow)$ the spin asymmetry, $r^*_F=\rho^*_F t_F$ the effective resistance, $t_F$ the thickness, and $\rho^*_F=(\rho_\uparrow+\rho_\downarrow)/4$ the effective resistivity of the ferromagnet, and $r_I=\rho_I t_I$ is the resistance of the interface.

\paragraph{Scattering theory.}

Since we are dealing with low-resistance metallic interfaces, the relevant resistances are those measured in the two-terminal setup, rather than the four-terminal resistances measured in a constriction or calculated within the Landauer-B\"uttiker approach. For spin-conserving interfaces the relation between the two is well-known \cite{Bauer}: the interface resistance appearing in series-resistor expressions is obtained from the Landauer-B\"uttiker resistance by subtracting the spurious contribution of the Sharvin resistance. The approach of Ref.\ \onlinecite{Bauer}, which takes into account the deviations of the distribution functions from equilibrium, can be readily applied to the periodic multilayer introduced above.

We use the result of Ref.\ \onlinecite{Bauer} for the two-terminal conductance $G^S$:
\begin{equation}
G^S = 2G_0 \sum_{ij\sigma\sigma'}[(I-T+R)^{-1}T]_{i\sigma,j\sigma'}
\end{equation}
where $i$, $j$ denote conduction channels, $G_0=e^2/h$, and the transmission and reflection matrices $T$ and $R$ are now $2\times2$ in spin space. The transmission and reflection matrices are calculated using the semiclassical concatenation rules \cite{Datta}. The irrelevant spin-flip scattering in the ferromagnetic layers is neglected, and the spin-diagonal transmission and reflection matrices across half of the ferromagnetic layer are written as
\begin{equation}
T^F_{i\sigma,j\sigma'}=\frac1{M_1}\frac{\delta_{\sigma\sigma'}}{1+s_\sigma}, \quad R^F_{i\sigma,j\sigma'}=\frac1{M_1}\frac{s_\sigma\delta_{\sigma\sigma'}}{1+s_\sigma}
\end{equation}
where $M_1$ is the number of conducting channels per spin in the adjacent normal metal, and $s_\sigma=r_\sigma M_1/2$, where $r_\sigma$ is the resistance of one spin channel (which includes the F/N interface resistance). The factor $\frac12$ comes from the fact that the supercell period contains half of the F layer at each edge. Concatenation of two such ``half-thick'' F layers leads to the correct scattering matrices for the whole F layer. The results of this calculation are identical to those of the circuit theory, Eqs.\ (6)-(7).

\paragraph{Circuit theory.}

A more general approach, not limited to periodic structures, is to employ the magnetoelectronic circuit theory \cite{Brataas.Nazarov.ea:PRL2000,Brataas.Nazarov.ea:EPJB2001,Brataas.PhysRep} extended to include spin-flip scattering \cite{supplement}. For an adjacent pair of layers L$_1$, L$_2$ in a magnetic multilayer, the charge $I^{0}$ and spin $\bar{I}^{s}$ currents in, say, layer L$_2$ are:
\begin{align}
I^{0}_{2}&=G\Delta f^{0}+\bar{G}^{s}\Delta \bar{f}^{s}-\bar{G}^{t}\cdot\bar{f}_1^{s}-\bar{G}^{r}\cdot\bar{f}_2^{s},\label{charge}\\
\bar{I}^{s}_{2}&=\bar{G}^{s} \Delta f^{0}+G\Delta \bar{f}^{s}-\hat{{\cal G}}^{t}\cdot\bar{f}_1^{s}-\hat{{\cal G}}^{r}\cdot\bar{f}_2^{s}.\label{spin}
\end{align}
Here $\Delta f^{0}=f^{0}_{1}-f^{0}_{2}$ and $\Delta f^{s}=f^{s}_{1}-f^{s}_{2}$ are interfacial drops of charge and spin components of the distribution function. We introduced $28$ parameters, including one scalar charge conductance $G$, three vector conductances $\bar{G}^{s}$, $\bar{G}^{t}$ and $\bar{G}^{r}$, and two tensor conductances $\hat{{\cal G}}^{t}$ and $\hat{{\cal G}}^{r}$ (see Supplemental Material \cite{supplement} for their definitions and relation to the notation used in Ref.\ \cite{Waintal.Myers.ea:Phys.Rev.B2000}). Equations (\ref{charge})-(\ref{spin}) represent the most general form of the boundary conditions; in particular, they include the effects of the mixing conductances, which are important in noncollinear magnetic multilayers \cite{Zwierzycki.Tserkovnyak.ea:PRB2005,Tserkovnyak.Brataas.ea:PRL2002,Kovalev.Bauer.ea:PRB2006}. They also reproduce the generalization of Valet-Fert theory to noncollinear systems \cite{Kovalev.Bauer.ea:PRB2002,Barnas:PRB2005}.

The expressions simplify for a non-magnetic, axially symmetric interface, for which $\bar{G}^{s}=\bar{G}^{t}=\bar{G}^{r}=0$, and the tensors $\hat{{\cal G}}^{t}$ and $\hat{{\cal G}}^{r}$ are diagonal in the axial reference frame. For highly transparent interfaces all conductances should be properly renormalized \cite{Schep.vanHoof.ea:PRB1997,Bauer.Tserkovnyak.ea:PRB2003}; the expressions are given in the Supplemental Material \cite{supplement}.

We apply the circuit theory to the FN$_1$(N$_2$N$_1$)$_\mathcal{N}$F spin valve, using Kirchhoff's rules for charge and spin conservation in each node. For simplicity, we assume that the spin accumulation is aligned parallel or perpendicular to the interface; the general case can be treated as a superposition of these alignments. Retaining only first-order terms in spin-flip scattering at each concatenation step, we find the magnetoresistance
\begin{align}
\Delta R = \frac{(\beta r^*_F)^2}{\tilde r_I m}\left[1- \frac{{\cal \tilde{G}}^t}{\tilde{G}}- (m^2-1)\frac{2{\cal \tilde{G}}^t+{\cal \tilde{G}}^r_1+{\cal \tilde{G}}^r_2}{6\tilde{G}}\right],\label{dr}
\end{align}
where the tilde accentuates the renormalized conductances \cite{supplement} for the given spin accumulation axis (for example, $2G_0/\tilde{G}= 2G_0/G-1/2M_1-1/2M_2$ \cite{Bauer}). Before renormalization,  $G=G_0(T_{\uparrow\uparrow}+T_{\downarrow\downarrow}+T_{\uparrow\downarrow}+T_{\downarrow\uparrow})$, ${\cal G}^{t}=2G_0(T_{\uparrow\downarrow}+T_{\downarrow\uparrow})$, and ${\cal G}^{r}_i=2G_0(R_{\uparrow\downarrow}^i+R_{\downarrow\uparrow}^i)$ corresponds to reflectance with incidence from metal N$_i$. When the number of layers is large, we can neglect $m$-independent terms and rewrite (\ref{dr}) as
\begin{equation}
\Delta R_{\parallel(\perp)} = \frac{(\beta r^*_F)^2}{\tilde r_I m}\left[1-\frac13 m^2\frac{{\cal G}^{sl}_{\parallel(\perp)}}{\tilde{G}}\right]
\label{mr}
\end{equation}
where $\tilde r_I = \tilde{G}^{-1}$ is the renormalized interface resistance, and we also introduced the spin-loss conductance ${\cal G}^{sl}={\cal G}^{t}+({\cal G}^{r}_1+{\cal G}^{r}_2)/2$. Note that ${\cal G}^{sl}$ does not need to be renormalized by the Sharvin resistance when calculated  up to the first order in the spin-flip processes.

To establish correspondence with the Valet-Fert model, we note that, to second order in $x$, we have $x/\sinh x\approx (1-x^2/6)$. Relating Eq.\ (\ref{mr}) and (\ref{drVF}), we find
\begin{equation}
\delta^2=2\frac{{\cal G}^{sl}}{\tilde G}\label{delta}
\end{equation}

The assumption of small $m\delta$ is, however, not essential.
Applying Eqs.~({\ref{charge}})-(\ref{spin}) to three contiguous non-magnetic layers \cite{supplement}, we find the following finite-difference equation for the spin accumulation:
\begin{equation}
{\cal D}^2 f^s_i=f^s_{i-1}-2f^s_i+f^s_{i+1},\label{eq:rec-1}
\end{equation}
where ${\cal D}^2=2 {\cal \tilde{G}}^{sl}/(\tilde{G}-{\cal \tilde{G}}^t)$. The most general solution of Eq.~(\ref{eq:rec-1})  has the form:
\begin{equation}
f^s_i=C_{1}e^{\delta i}+C_{2}e^{-\delta i},\label{eq:sol}
\end{equation}
where $\delta=\ln \left\{ 1+({\cal D}^2/2)[1+(1+4/{\cal D}^2)^{1/2}]\right\}$.
This is identical to the solution of the Valet-Fert equations \cite{VF1993} and generalizes the definition of $\delta$ (\ref{delta}) to the strong spin-flip scattering case.
If the spin-flip scattering is weak, we recover Eq.~(\ref{delta}), since in this limit $\delta\approx {\cal D}$.

Equation (\ref{delta}) shows that $\delta$ is proportional not to the spin-flip scattering probability at the interface (as it has been usually assumed \cite{Bass}), but to its square root. Thus, for example, a seemingly large value $\delta\approx0.24$ deduced experimentally for the Cu/Pd interface corresponds to a spin-flip scattering probability of less than 2\%.

For weak spin-flip scattering, the parameter $\delta$ measured in multilayer ($m\gg1$) magnetoresistance experiments depends only on the sum of spin-flip transmission ($T_{\uparrow\downarrow}$) and reflection ($R^i_{\uparrow\downarrow}$) probabilities. These parameters are not related through unitarity, and there is no reason to assume any specific relation between them for a thin interface. In fact, spin transport in circuits containing spin-non-conserving interfaces generally depends separately on these probabilities. Therefore, the parameter $\delta$ and the area-resistance product of the interface do not provide complete information needed for the description of arbitrary magnetoelectronic circuits.

We also note that the $T^{(m)}_{\uparrow\downarrow}$ and $R^{(m)}_{\uparrow\downarrow}$ components of the matrices, which are obtained by concatenating $m$ identical spin-non-conserving scattering matrices, converge with each other when $m$ becomes large: $T^{(m)}_{\uparrow\downarrow}\approx R^{(m)}_{\uparrow\downarrow}\approx m(T_{\uparrow\downarrow}+R_{\uparrow\downarrow})$. (The latter equality holds as long as $T^{(m)}_{\uparrow\downarrow}\ll T^{(m)}_{\uparrow\uparrow}$.) For this reason, the resistance and parameter $\delta=t/l_{sf}$ completely describe the behavior of a sufficiently thick non-magnetic \emph{bulk} layer in an arbitrary circuit, as assumed in the Valet-Fert theory.

\paragraph{First-principles calculations.}

The spin-resolved transmittances and reflectances were calculated using the Landauer-B\"uttiker approach \cite{Datta} implemented within the tight-binding linear muffin-tin orbital (TB-LMTO) method \cite{Turek}. The discretized representation was used for the coordinate operator in transport calculations \cite{Kudr}, and SOC was included as a perturbation to the LMTO potential parameters \cite{Belashchenko-SOC,Turek-SOC}. The generalized gradient approximation is used for exchange and correlation \cite{PBE}.

We focus on the Cu/Pd interface, for which the experimental measurements yield a fairly large parameter $\delta\approx0.24$, with relatively narrow error bars \cite{Kurt}. We consider (111) and (001) interface orientations, with the spin quantization axis, corresponding to the polarization of the spin current in a device, aligned either parallel or perpendicular to the interface. We assume that the atomic positions lie on the ideal face-centered cubic lattice with a lattice constant $a=3.818$ \AA. In addition to the ideal interfaces, several simple intermixing models are considered for the (111) orientation.

Some care needs to be taken to define the spin-flip scattering probabilities, bearing in mind that, owing to the presence of SOC in the bulk, the electronic states in each spin reservoir are already not pure spin-up and spin-down spinors. This bulk spin mixing should be separated from the spin-flip scattering at the interface.

To define the spin-resolved interfacial transmittance $T_{\sigma\sigma'}$ and reflectance $R^i_{\sigma\sigma'}$ (where $i=\mathrm{Cu}$ or Pd), we turn off SOC in the leads and introduce ``ramp-up'' regions where SOC is gradually increased as one moves away from the embedding planes toward the Cu/Pd interface. For generic $\mathbf{k}$-points this ``adiabatic embedding'' allows pure spin states in the leads to evolve without scattering into the bulk eigenstates, and the spin-dependent scattering probabilities are thus properly defined \cite{note-Pd}. An exception occurs near the boundaries of the projections of the Fermi sheets, where the group velocity is nearly parallel to the interface. Here the deformation of the Fermi surface by SOC can lead to strong reflection.

\begin{figure}[htb]
\includegraphics[width=0.46\columnwidth]{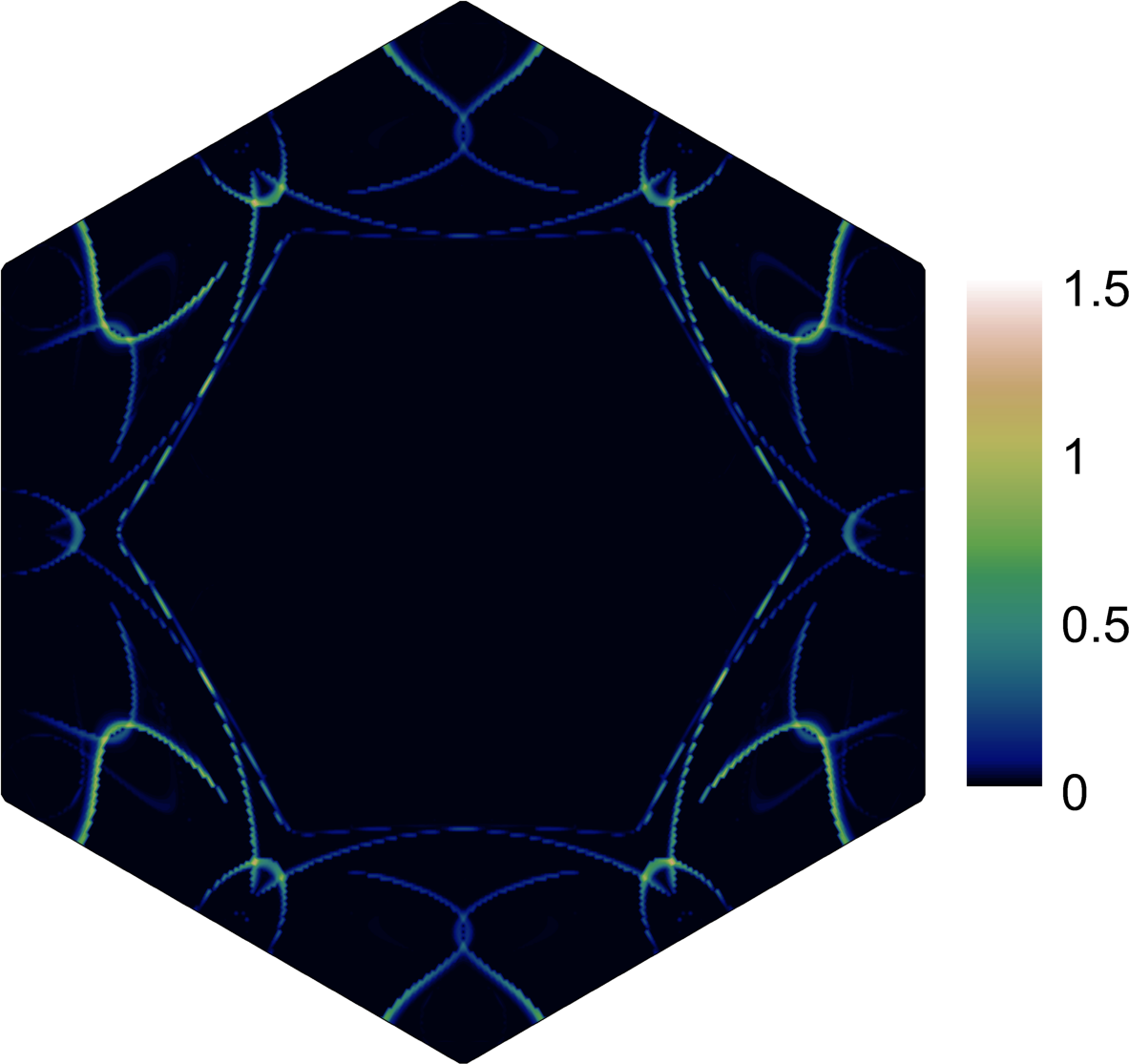}
\caption{$\mathbf{k}$-resolved spin-flip reflectance $R_{\uparrow\downarrow}$ for the test Pd system, in which SOC is gradually suppressed away from a (111) plane. The spin quantization axis points up, parallel to the interface.}
\label{pdpd}
\end{figure}

To examine the effect of adiabatic embedding on the Pd side, we consider a Pd slab of thickness $D$, located at $|x|<D/2$ and attached to Pd leads without SOC at $|x|>D/2$, with the SOC parameters scaled by a function $f(|x|)$ such that $f(0)=1$ and $f(D/2)=0$. We used a simple trapezoidal form of $f(x)$, which is constant over a few atomic layers near the interface and then declines linearly to zero; the results are insensitive to the shape of $f(x)$. As long as $D$ is at least a few dozen monolayers in this test system, $T_{\uparrow\downarrow}$ is negligible, while $R_{\uparrow\downarrow}$ is 2--4 times smaller compared to $R^\mathrm{Pd}_{\uparrow\downarrow}$ in the Cu/Pd system with a similar ramp-up region on the Pd side. Fig.\ \ref{pdpd} shows that the $\mathbf{k}$-resolved $R_{\uparrow\downarrow}$ in the test system is indeed significant only near the edges of the Fermi surface projections. As expected, $R_{\uparrow\downarrow}$ in the test Pd system quickly saturates as the width $D$ is increased. Qualitatively, the situation is analogous to the ballistic scattering from a ferromagnetic domain wall \cite{Brataas1999}.

Strong reflection near the edges of the Fermi surface projection persists in the Cu/Pd system with adiabatic embedding. Since these edges are in no way special for the scattering from the abrupt Cu/Pd interface, it should be attributed to the reflection from the ramp-up region. Therefore, we subtract $R_{\uparrow\downarrow}$ for the test Pd system from $R^\mathrm{Pd}_{\uparrow\downarrow}$ for the Cu/Pd interface. Since the former is a few times smaller than the latter, the uncertainties inherent in this procedure lead to relatively small errors in $\delta$ compared to the experimental uncertainty \cite{note-filtering}.

In addition to ideal (111) and (001) interfaces, we considered several simple models of roughness with intermixing in one monolayer for the (111) interface, with the following structures of this monolayer: (A) 1:1 superlattice (50/50 model), (B) $2\times2$ ordering of Pd atoms within the Cu monolayer (75/25 model), (C) $2\times2$ ordering of Cu atoms within the Pd monolayer (25/75 model).

The results are listed in Table \ref{data}. Here $\bar R^\mathrm{Cu}_{\uparrow\downarrow}/A$ and $\bar R^\mathrm{Pd}_{\uparrow\downarrow}/A$ are the specific spin-flip reflectances for Cu with SOC embedded in Cu without SOC, and for adiabatically embedded Pd with SOC, respectively. The integration is performed using a mesh of $256\times256$ points in the full two-dimensional Brillouin zone; a coarser $64\times64$ mesh yields very similar results. For each interface we consider two orientations of the spin quantization axis, parallel ($\parallel$) and perpendicular ($\perp$) to the interface, which reflects the orientation of the spin accumulation in the device. In the parallel case we average $T_{\uparrow\downarrow}$ and $R^s_{\uparrow\downarrow}$ over two orthogonal in-plane orientations of the spin quantization axis; we also average over the reversed spin indices, e.g., $T_{\uparrow\downarrow}$ and $T_{\downarrow\uparrow}$, as well $T_{\uparrow\uparrow}$ and $T_{\downarrow\downarrow}$. The deviations from axial symmetry are appreciable only for the 50/50 model of the (111) interface, where they reach 35\% for $R^\mathrm{Cu}_{\uparrow\downarrow}$.

\begin{table*}[htb]
\caption{Spin-dependent scattering at the Cu/Pd interfaces. Conductances per area are in PS/m$^2$; $2AR$ in f$\Omega\cdot$m$^2$.}
\begin{tabular}{|c|c|c|c|c|c|c|c|c|c|c|c|c|}
\hline
Plane & Structure & $\mathbf{M}$ & $G_0T_{\uparrow\uparrow}/A$ & $G_0T_{\uparrow\downarrow}/A$ & $G_0R^\mathrm{Cu}_{\uparrow\downarrow}/A$ & $G_0R^\mathrm{Pd}_{\uparrow\downarrow}/A$ & $G_0\bar R^\mathrm{Cu}_{\uparrow\downarrow}/A$ & $G_0\bar R^\mathrm{Pd}_{\uparrow\downarrow}/A$ & ${\cal G}^{sl}/A$ & $\tilde G/(2A)$ & 2$AR$ & $\delta$ \\
\hline
\multirow{2}{*}{$(001)$} & \multirow{2}{*}{Ideal} & $\parallel$ &0.30&0.003&0.016&0.033&0.0005&0.013&0.083&0.59&1.70& 0.38 \\
                         &                        & $\perp$     &0.30&0.003&0.031&0.040&0.0007&0.017&0.119&0.59&1.70& 0.45 \\
\hline
\multirow{8}{*}{$(111)$} & \multirow{2}{*}{Ideal} & $\parallel$ &0.32&0.008&0.010&0.039&0.0003&0.010&0.108&0.70&1.43& 0.39 \\
                         &                        & $\perp$     &0.32&0.011&0.017&0.052&0.0004&0.019&0.145&0.70&1.43& 0.45 \\
\cline{2-13}
                         & \multirow{2}{*}{50/50} & $\parallel$ &0.31&0.009&0.011&0.044&0.0003&0.010&0.125&0.66&1.51& 0.43 \\
                         &                        & $\perp$     &0.31&0.012&0.020&0.061&0.0004&0.019&0.173&0.66&1.51& 0.51 \\
\cline{2-13}
                         & \multirow{2}{*}{75/25} & $\parallel$ &0.31&0.010&0.011&0.048&0.0003&0.010&0.137&0.65&1.53& 0.46 \\
                         &                        & $\perp$     &0.31&0.014&0.020&0.067&0.0004&0.019&0.192&0.65&1.53& 0.54 \\
\cline{2-13}
                         & \multirow{2}{*}{25/75} & $\parallel$ &0.32&0.010&0.011&0.049&0.0003&0.010&0.141&0.71&1.41& 0.45 \\
                         &                        & $\perp$     &0.32&0.014&0.019&0.066&0.0004&0.019&0.188&0.71&1.41& 0.52 \\

\hline
\end{tabular}
\label{data}
\end{table*}

In all cases listed in Table \ref{data} the spin-loss conductance ${\cal G}^{sl}$ is dominated by spin-flip reflection. Thus, the parameter $\delta$ is not directly related to the probability of a spin flip in transmission, as it has been previously assumed \cite{Bass}.

\begin{figure}[htb]
\includegraphics[width=0.85\columnwidth]{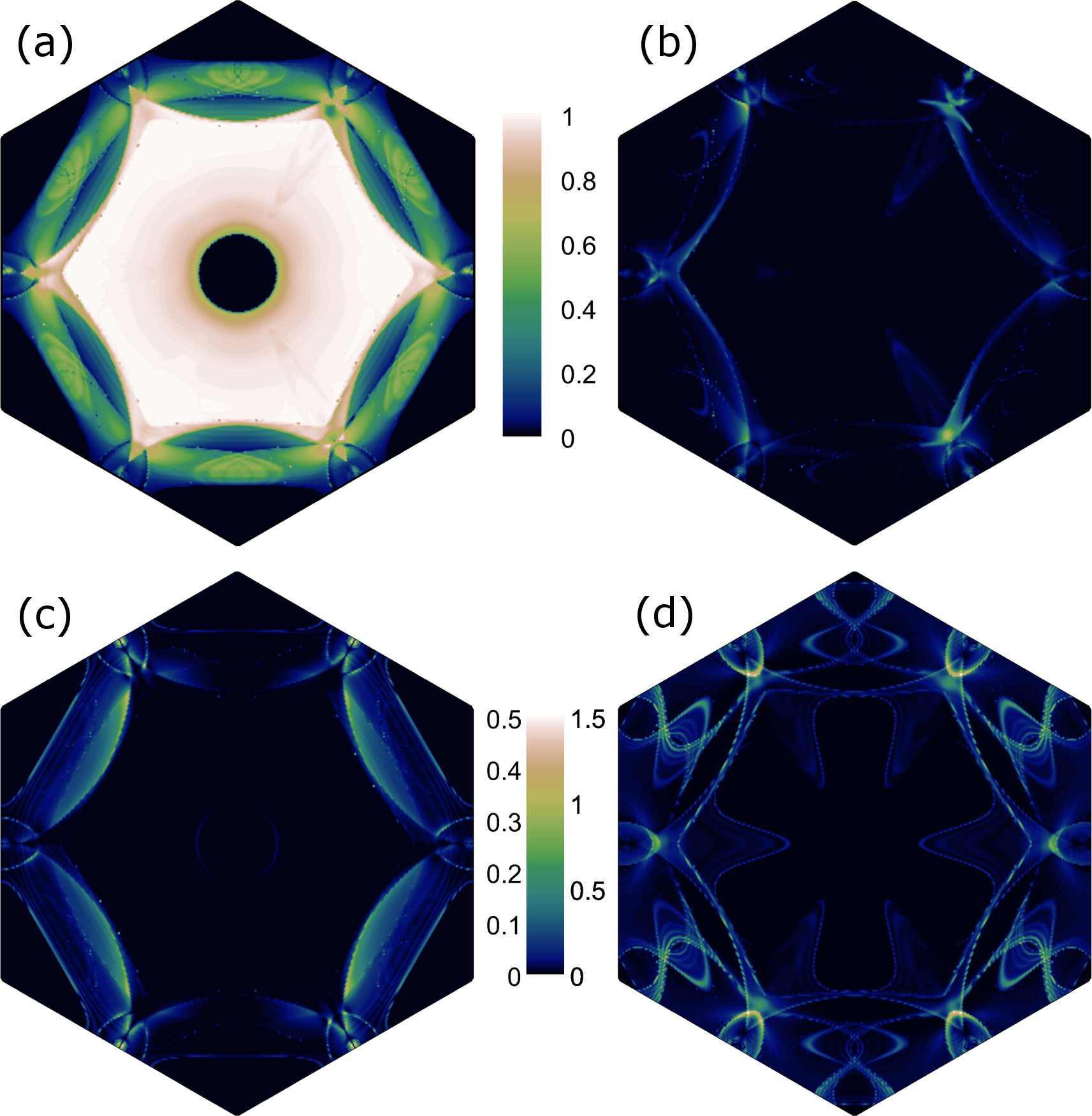}
\caption{$\mathbf{k}$-resolved transmittances $T_{\sigma\sigma'}$ and reflectances $R^s_{\sigma\sigma'}$ for the Cu/Pd (111) interface. (a) $T_{\uparrow\uparrow}$, (b) $T_{\uparrow\downarrow}$, (c) $R^\mathrm{Cu}_{\uparrow\downarrow}$, (d) $R^\mathrm{Pd}_{\uparrow\downarrow}$. The spin quantization axis points up, parallel to the interface.}
\label{cupd}
\end{figure}

Fig.\ \ref{cupd} shows $\mathbf{k}$-resolved transmittances and reflectances for the (111) interface with magnetization parallel to the interface. Note the mirror symmetry in the plane perpendicular to the spin quantization axis. Fig.\ \ref{cupd}(d) shows strong reflection at the Fermi edges, similar to Fig.\ \ref{pdpd}, which is due to the adiabatic embedding on the Pd side. However, contrary to Fig.\ \ref{pdpd}, significant spin-flip reflection is also seen at generic $\mathbf{k}$-points in Fig.\ \ref{cupd}(d), which originates at the Cu/Pd interface.

The values of the parameter $\delta$ for devices with in-plane ($\parallel$) spin accumulation (Table \ref{data}) can be directly compared with the experimental value $\delta=0.24^{+0.06}_{-0.03}$ \cite{Kurt}. The results for (001) and (111) interface orientations are quite similar and in reasonable agreement with experiment. In agreement with Ref.\ \onlinecite{Galinon}, the calculated interface area-resistance product $AR$ is overestimated by 65-100\% and is not strongly affected by intermixing. Intermixing also has a relatively small effect on $\delta$, increasing it by a small amount. Due to the fairly large size mismatch, the structure of the Cu/Pd multilayer can exhibit significant disorder and strain relaxation, which may lead to the discrepancy in the area-resistance product. The overestimation of $\delta$ may be due to the same reason.

Table \ref{data} shows that $\delta$ becomes notably larger when the spin accumulation is oriented perpendicular to the interface. This angular dependence can be tested in experiments on multilayers \cite{Bass,Bass2015} by utilizing ferromagnetic layers with perpendicular magnetization. Anisotropy of a similar kind was found for the spin relaxation rate in thin films \cite{Long1,Long2,Long3}. This spin relaxation is due to spin-flip reflection at the film surface, and it can also be described using the generalized circuit theory.

In conclusion, we have formulated a theory of spin loss at metallic interfaces, linking the calculable spin-dependent scattering properties of an interface with the phenomenological parameter $\delta$ measured in experiments on magnetoresistance in multilayers. This relation [Eq. (\ref{delta})] shows that spin-flip scattering on the order of a few percent yields $\delta$ that is comparable to unity. First-principles calculations for the Cu/Pd interface give $\delta$ in reasonable agreement with experiment, but somewhat overestimated. Understanding of spin loss at metallic interfaces is an important ingredient for the analysis of spin transport in magnetic heterostructures with strong spin-orbit coupling.

\begin{acknowledgments}

AK is much indebted to Gerrit Bauer for stimulating discussions on the circuit theory with spin-flip scattering. This work was supported by the National Science Foundation through Grant No.\ DMR-1308751 and the Nebraska MRSEC, Grant No.\ DMR-1420645, as well as by the DOE Early Career Award {DE-SC0014189} (AK) and the EPSRC CCP9 Flagship project, EP/M011631/1 (MvS). The computations were performed utilizing the Holland Computing Center of the University of Nebraska.

\end{acknowledgments}

\balancecolsandclearpage

\onecolumngrid

\centerline{\large\bf Supplemental Material}

\setcounter{figure}{0}
\setcounter{equation}{0}
\makeatletter
\renewcommand{\thefigure}{S\arabic{figure}}
\renewcommand{\theequation}{S\arabic{equation}}
\renewcommand{\bibnumfmt}[1]{[S#1]}
\renewcommand{\citenumfont}[1]{S#1}

\section{Circuit theory in the presence of spin-flip scattering}

Consider two metallic nodes separated by a scattering region. The current in each node depends on the potential drop and on the spin accumulation drop between the nodes. The current evaluated for node $2$ is \cite{S_Brataas.Nazarov.ea:EPJB2001}
\begin{equation}
\hat{I}_2=G_0\sum_{nm}\left[\hat{t}^{\prime}_{mn}\hat{f}_{1}(\hat{t}^{\prime}_{nm})^{\dagger}-\left(M_2\hat{f}_{2}-\hat{r}_{mn}\hat{f}_{2}(\hat{r}_{nm})^{\dagger}\right)\right],\label{current}
\end{equation}
where $G_0=e^2/h$, $\hat{r}_{mn}$ is the spin-dependent reflection amplitude for electrons reflected from channel $n$ into channel $m$ in node 2, and $\hat{t}^{\prime}_{mn}$ is the spin-dependent transmission amplitude for electrons transmitted from channel $n$ in node 1 into channel $m$ in node 2. Note that the ensuing results can be easily rewritten for the current $\hat{I}_{1}$ in node $1$. Spin-flip scattering at the interface makes the matrices $\hat{r}_{mn}$ and $\hat{t}^{\prime}_{mn}$ non-diagonal in spin space.

Let us introduce a matrix:
\begin{equation}
\check{S}_{mn}=\left(\begin{array}{cc}
\hat{r}_{mn} & \hat{t}'_{mn}\\
\hat{t}_{mn} & \hat{r}'_{mn}
\end{array}\right),
\end{equation}
where $\hat r'$ and $\hat t$ are the amplitudes of reflection and transmission into node 1.
Charge conservation requires $\check{S}\check{S}^{\dagger}=\check{1}$, and, therefore,
\begin{equation}
\sum_{mn}\check{S}_{mn}\check{S}_{mn}^{\dagger}=\check{M}=\hat \sigma^0\otimes \hat M,\label{Smatrix}
\end{equation}
where $\hat \sigma^0$ is a unit matrix in spin space, the symbol $\otimes$ denotes the Kronecker product, and $\hat M$ is a diagonal matrix with elements $M_{ii}=M_i$ representing the number of channels in electrode $i$.
We extract only part of Eq. (\ref{Smatrix})
that contains $\hat{r}_{mn}$ and $\hat{t}'_{mn}$ coefficients:
\begin{equation}
\sum_{mn}\hat{r}_{mn}\hat{r}_{mn}^{\dagger}+\hat{t}'_{mn}(\hat{t}'_{mn})^{\dagger}=M_2\hat\sigma^0\label{constr}
\end{equation}
leading to three independent constraints on the elements of the $S$ matrix. If the system has time reversal symmetry, the total $S$ matrix also satisfies $S=S^{T}$.

The spin-dependent distribution functions in nodes $1$ and $2$, as well as the current matrix, can be expressed via the Pauli matrices $\hat{\sigma}^{1}$, $\hat{\sigma}^{2}$, $\hat{\sigma}^{3}$ and the unit matrix $\hat{\sigma}^{0}$: $\hat{f}_{1}=\hat\sigma^0f_{1}^{0}+\hat{\sigma}\bar{f}_{1}^{s}$,
$\hat{f}_{2}=\hat\sigma^0f_{2}^{0}+\hat{\sigma}\bar{f}_{2}^{s}$, $\hat{I}=(\hat\sigma^0I^{0}+\hat{\sigma}\bar{I}^{s})/2$.
We express the scattering amplitudes with the help of notations proposed
in Ref. \cite{S_Waintal.Myers.ea:Phys.Rev.B2000}. Denoting the unit matrix as $\hat\sigma^0$, we define $\mathcal{R}_{mn}^{\mu\nu}=\mbox{Tr}[(\hat{r}_{mn}\otimes\hat{r}_{mn}^{*})\cdot(\hat{\sigma}^{\mu}\otimes\hat{\sigma}^{\nu})]/4$
and $\mathcal{T}_{mn}^{\mu\nu}=\mbox{Tr}[(\hat{t}'_{mn}\otimes\hat{t}^{\prime*}_{mn})\cdot(\hat{\sigma}^{\mu}\otimes\hat{\sigma}^{\nu})]/4$.

The circuit theory expression (\ref{current}) can now be rewritten in the form of Eqs.\ (4)-(5) of the main text, with the following definitions of the conductances:
\begin{align}
&G=2G_0\sum_{mn}\mathcal{T}_{mn}^{\nu\nu},\quad
G_{i}^{s}=2G_0\sum_{mn}(\mathcal{T}_{mn}^{i0}+\mathcal{T}_{mn}^{0i}+i\varepsilon_{ijk}\mathcal{T}_{mn}^{jk}),\\
&G_{i}^{t}=4G_0\sum_{mn}i\varepsilon_{ijk}\mathcal{T}_{mn}^{jk},\quad
G_{i}^{r}=4G_0\sum_{mn}i\varepsilon_{ijk}\mathcal{R}_{mn}^{jk},\\
&\mathcal{G}_{ij}^{t}=2G_0\delta_{ij}^{kl}\sum_{mn}(\mathcal{T}_{mn}^{kl}+\mathcal{T}_{mn}^{lk}+i\varepsilon_{klv}[\mathcal{T}_{mn}^{0v}-\mathcal{T}_{mn}^{v0}]),\\
&\mathcal{G}_{ij}^{r}=2G_0\delta_{ij}^{kl}\sum_{mn}(\mathcal{R}_{mn}^{kl}+\mathcal{R}_{mn}^{lk}+i\varepsilon_{klv}[\mathcal{R}_{mn}^{0v}-\mathcal{R}_{mn}^{v0}]),
\end{align}
where $\delta_{ij}^{kl}=\delta_{ik}\delta_{jl}-\delta_{ij}\delta_{kl}$, and summation over the repeated indices is assumed everywhere.

In the case of a non-magnetic (disordered) interface with axial symmetry, $\bar{G}^{s}=\bar{G}^{t}=\bar{G}^{r}=0$, while the tensors $\hat{{\cal G}}^{t}$ and $\hat{{\cal G}}^{r}$ are diagonal in the reference frame aligned with the symmetry axis. These simplifications lead to the following expressions for the currents in the nodes:
\begin{align}
I_{0}&=G\Delta f_{0},\label{charge-1}\\
\bar{I}_{2}^{s}&=(G-\mathcal{G}^{t})\Delta\bar{f}_{s}-\mathcal{G}_{2}^{sl}\bar{f}^s_2,\label{spin-1}\\
\bar{I}_{1}^{s}&=(G-\mathcal{G}^{t})\Delta\bar{f}_{s}+\mathcal{G}_{1}^{sl}\bar{f}^s_2,\label{spin-2}
\end{align}
where we introduced the spin-loss conductance $\mathcal{G}_{1(2)}^{sl}=\mathcal{G}_{1(2)}^{r}+\mathcal{G}^{t}$
calculated along one of the symmetry axes in the nodes.

\section{Renormalizations for Ohmic contacts}

It is well known that interface resistances in transparent Ohmic contacts are renormalized by the Sharvin resistance \cite{S_Schep.vanHoof.ea:PRB1997,S_Bauer.Tserkovnyak.ea:PRB2003}. The circuit theory can be generalized to account for the drift contributions in the nodes by renormalizing the conductances $G$, $\mathcal{G}^{t}$, and $\mathcal{G}_{1(2)}^{sl}$. This can be done by connecting nodes $1$ and $2$ to proper reservoirs with spin-dependent distribution functions $\hat{f}_{L}$ and $\hat{f}_{R}$ via transparent contacts. The currents in the nodes then become $\hat{I}_{1}=2G_0M_{1}(\hat{f}_{L}-\hat{f}_{1})$ and $\hat{I}_{2}=2G_0M_{2}(\hat{f}_{2}-\hat{f}_{R})$, where $M_{1(2)}$ describe the number of channels in the nodes. Substituting these currents in Eqs.\ (\ref{charge-1}), (\ref{spin-1}), and (\ref{spin-2}), we arrive at the amended circuit theory:
\begin{align}
I_{0}&=G(\Delta f_{0}+\dfrac{I_{0}}{4G_0M_{1}}+\dfrac{I_{0}}{4G_0M_{2}}),\\
I^s_1&=(G-\mathcal{G}^{t})(\Delta f_{s}+\dfrac{I^s_1}{4G_0M_{1}}+\dfrac{I^s_2}{4G_0M_{2}})+\mathcal{G}_{1}^{sl}(f_{s}^{1}+\dfrac{I^s_1}{4G_0M_{1}}),\\
I^s_2&=(G-\mathcal{G}^{t})(\Delta f_{s}+\dfrac{I^s_1}{4G_0M_{1}}+\dfrac{I^s_2}{4G_0M_{2}})-\mathcal{G}_{2}^{sl}(f_{s}^{2}-\dfrac{I^s_2}{4G_0M_{2}}).
\end{align}
These equations are equivalent to Eqs. (\ref{charge-1})-(\ref{spin-2}) after the substitution $G\rightarrow\tilde{G}$,
$\mathcal{G}^{t}\rightarrow\tilde{\mathcal{G}}^{t}$, and $\mathcal{G}_{1(2)}^{sl}\rightarrow\tilde{\mathcal{G}}_{1(2)}^{sl}$,
where
\begin{align}
&\dfrac{2}{\tilde{G}}=\dfrac{2}{G}-\dfrac{1}{2G_0M_{1}}-\dfrac{1}{2G_0M_{2}}, \\
&\dfrac{2}{\tilde{G}-\tilde{\mathcal{G}}^{t}+\frac{\tilde{\mathcal{G}}_{1}^{sl}\tilde{\mathcal{G}}_{2}^{sl}}{\tilde{\mathcal{G}}_{1}^{sl}+\tilde{\mathcal{G}}_{2}^{sl}}}
=\dfrac{2}{G-\mathcal{G}^{t}+\frac{\mathcal{G}_{1}^{sl}\mathcal{G}_{2}^{sl}}{\mathcal{G}_{1}^{sl}+\mathcal{G}_{2}^{sl}}}-\dfrac{1}{2G_0M_{1}}-\dfrac{1}{2G_0M_{2}},\\
&\dfrac{1}{\tilde{\mathcal{G}}_{1}^{sl}}
=\dfrac{1}{\mathcal{G}_{1}^{sl}}-\dfrac{1}{2G_0M_{1}}-\dfrac{\mathcal{G}_{2}^{sl}/\mathcal{G}_{1}^{sl}-M_{2}/M_{1}}{\mathcal{G}_{1}^{sl}+\mathcal{G}_{2}^{sl}+2\mathcal{G}_{1}^{sl}\dfrac{\mathcal{G}_{2}^{sl}-2G_0M_{2}}{G-\mathcal{G}^{t}}},\\
&\dfrac{1}{\tilde{\mathcal{G}}_{2}^{sl}}
=\dfrac{1}{\mathcal{G}_{2}^{sl}}-\dfrac{1}{2G_0M_{2}}-\dfrac{\mathcal{G}_{1}^{sl}/\mathcal{G}_{2}^{sl}-M_{1}/M_{2}}{\mathcal{G}_{1}^{sl}+\mathcal{G}_{2}^{sl}+2\mathcal{G}_{2}^{sl}\dfrac{\mathcal{G}_{1}^{sl}-2G_0M_{1}}{G-\mathcal{G}^{t}}}.
\end{align}
Note that these equations can be further simplified in the symmetric case, $\mathcal{G}_{1}^{sl}=\mathcal{G}_{2}^{sl}$
and $M_{1}=M_{2}$.

\section{Transport in N$_1|$N$_2$ superlattice}

We now assume that we have a superlattice constructed out of repeated interfaces between two normal metals N$_1$ and N$_2$. We take nodes in both N$_1$ and N$_2$ layers, and the conductances $\tilde{G}$, $\tilde{\mathcal{G}}^{t}$, $\tilde{\mathcal{G}}_{1}^{sl}$, and $\tilde{\mathcal{G}}_{2}^{sl}$
describe the two nodes. We arrive at the following equations for the spin current in node $i$:
\begin{align}
I^s_i&=(\tilde{G}-\tilde{\mathcal{G}}^{t})(f^s_{i-1}-f^s_{i})-\tilde{\mathcal{G}}_{1}^{sl}f^s_i,\label{SL1-1}\\
I^s_i&=(\tilde{G}-\tilde{\mathcal{G}}^{t})(f^s_{i}-f^s_{i+1})+\tilde{\mathcal{G}}_{2}^{sl}f^s_i,\label{SL2-1}
\end{align}
which leads to the recursive formula:
\begin{equation}
\dfrac{2\tilde{\mathcal{G}}^{sl}}{\tilde{G}-\tilde{\mathcal{G}}^{t}}f^s_i=f^s_{i-1}-2f^s_{i}+f^s_{i+1},\label{eq:rec_app}
\end{equation}
where $\tilde{\mathcal{G}}^{sl}=(\tilde{\mathcal{G}}_{1}^{sl}+\tilde{\mathcal{G}}_{2}^{sl})/2$.
This recursive equation has the following solution:
\begin{equation}
f_{s}^{i}=C_{1}e^{\delta i}+C_{2}e^{-\delta i},\label{eq:sol-1}
\end{equation}
where
\begin{equation}
\delta=\ln\left[1+\dfrac{\tilde{\mathcal{G}}^{sl}}{\tilde{G}-\tilde{\mathcal{G}}^{t}}\left(1+\sqrt{1+\dfrac{2(\tilde{G}-\tilde{\mathcal{G}}^{t})}{\tilde{\mathcal{G}}^{sl}}}\right)\right],
\end{equation}
and the constants $C_{1}$ and $C_{2}$ depend on the boundary conditions.

\section{Accounting for the bulk contribution}

\begin{figure}[htb]
\includegraphics[width=0.35\textwidth]{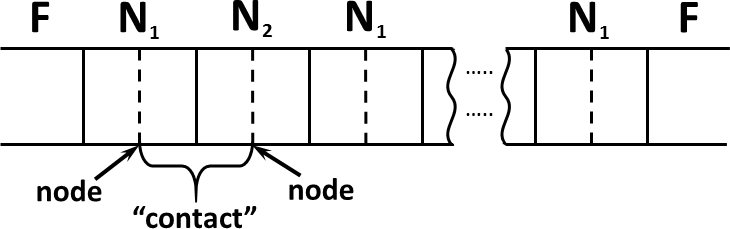}
\caption{Partitioning of the multilayer in nodes and contacts.}
\label{contact}
\end{figure}

Within the circuit theory, spin transport across a non-magnetic interface that is axially symmetric (either microscopically or after averaging over crystallite orientations) is fully characterized by four conductances: $\tilde{G}$, $\tilde{\mathcal{G}}^{t}$, $\tilde{\mathcal{G}}_{1}^{sl}$, and $\tilde{\mathcal{G}}_{2}^{sl}$. We will also refer to the quantities $\tilde{\mathcal{G}}^{s}=\tilde{G}-\tilde{\mathcal{G}}^{t}$, which appear in Eqs.\ (\ref{SL1-1})-(\ref{SL2-1}), as spin conductances. In the main text of the paper we have neglected the resistivities of the bulk metallic layers and assumed that spin relaxation occurs only at the interfaces, in order to simplify the resulting expressions. These features can be restored by placing the circuit nodes in the middle of the bulk layers. A contact between two nodes is then defined to include both the physical interface and the adjacent bulk regions extending up to these nodes, as shown in Fig.\ \ref{contact}.

Spin-transport in a bulk diffusive region $i$ is assumed to obey the Valet-Fert model, which yields $\tilde{\mathcal{G}}_{bi}^{s}=\tilde{G}_{bi}\delta_{i}/\sinh\delta_{i}$ and $\tilde{\mathcal{G}}_{bi}^{sl}=\tilde{G}_{bi}\delta_{i}\tanh(\delta_{i}/2)$, where $\delta_{i}=t_i/l^i_{sf}$ is defined similar to the spin-memory loss parameter for an interface. We have added a subscript $b$ to distinguish bulk and interface conductances in the following. There is only one $\tilde{\mathcal{G}}_{bi}^{sl}$ parameter, because a bulk region is left-right symmetric. Thus, two parameters $\tilde{G}_{bi}$ and $\delta_i$ completely describe a diffusive bulk layer. (A general interface can not be fully described in this way, because four independent conductances can not be reduced to two parameters $\tilde{G}$ and $\delta$.)

Introducing the conductances $\tilde{G}_a$, $\tilde{\mathcal{G}}_a^{s}$, $\tilde{\mathcal{G}}_{a1}^{sl}$, and $\tilde{\mathcal{G}}_{a2}^{sl}$ for the composite three-layer ``contact,'' we can apply Eq.\ (9) from the main text to obtain
\begin{equation}
{\cal D}^2= \frac{\tilde{\mathcal{G}}_{a1}^{sl}+\tilde{\mathcal{G}}_{a2}^{sl}}{\tilde{\mathcal{G}}^{s}_a},\label{eq:new}
\end{equation}
which now fully takes into account the bulk contributions. The composite conductances can be obtained by concatenating the interface with the adjacent bulk regions using the circuit theory:

\begin{align}
\tilde{\mathcal{G}}_a^{s}&=\frac{ \tilde{\mathcal{G}}_{b1}^{s} \tilde{\mathcal{G}}_{b2}^{s} \tilde{\mathcal{G}}^{s}}
{(\tilde{\mathcal{G}}_{b1}^{s} + \tilde{\mathcal{G}}_{c1}^{sl}) (\tilde{\mathcal{G}}_{b2}^{s} + \tilde{\mathcal{G}}_{c2}^{sl})
+(\tilde{\mathcal{G}}_{b1}^{s} + \tilde{\mathcal{G}}_{b2}^{s} + \tilde{\mathcal{G}}_{c1}^{sl} + \tilde{\mathcal{G}}_{c2}^{sl}) \tilde{\mathcal{G}}^{s}},\\
\tilde{\mathcal{G}}_{a1}^{sl}&=
\frac{ (\tilde{\mathcal{G}}_{b1}^{s} + \tilde{\mathcal{G}}_{c1}^{sl}) [\tilde{\mathcal{G}}_{b2}^{s} (\tilde{\mathcal{G}}_{c2}^{sl} + \tilde{\mathcal{G}}_{b2}^{sl})+\tilde{\mathcal{G}}_{b2}^{sl} \tilde{\mathcal{G}}_{c2}^{sl}] + [(\tilde{\mathcal{G}}_{b1}^{s} + \tilde{\mathcal{G}}_{c1}^{sl} + \tilde{\mathcal{G}}_{c2}^{sl})\tilde{\mathcal{G}}_{b2}^{sl}  +
    \tilde{\mathcal{G}}_{b2}^{s} (\tilde{\mathcal{G}}_{c1}^{sl} + \tilde{\mathcal{G}}_{c2}^{sl} + \tilde{\mathcal{G}}_{b2}^{sl})] \tilde{\mathcal{G}}^{s} }
    {(\tilde{\mathcal{G}}_{b1}^{s} + \tilde{\mathcal{G}}_{c1}^{sl}) (\tilde{\mathcal{G}}_{b2}^{s} + \tilde{\mathcal{G}}_{c2}^{sl}) + (\tilde{\mathcal{G}}_{b1}^{s} +
    \tilde{\mathcal{G}}_{b2}^{s} + \tilde{\mathcal{G}}_{c1}^{sl} + \tilde{\mathcal{G}}_{c2}^{sl}) \tilde{\mathcal{G}}^{s}},
\end{align}
\normalsize
where $\tilde{\mathcal{G}}^{sl}_{ci}=\tilde{\mathcal{G}}^{sl}_{bi}+\tilde{\mathcal{G}}^{sl}_{i}$. The expression for $\tilde{\mathcal{G}}_{a2}^{sl}$ is obtained from $\tilde{\mathcal{G}}_{a1}^{sl}$ by interchanging the indices 1 and 2. We also have $\tilde{G}^{-1}_a=\tilde{G}^{-1}+\tilde{G}_{b1}^{-1}+\tilde{G}_{b2}^{-1}$.

Expanding of Eq.~(\ref{eq:new}) to first order in spin-flip scattering results in
\begin{equation}
{\cal D}^2\approx \frac{\mathcal{G}_{1}^{sl}+\mathcal{G}_{2}^{sl}+2\mathcal{G}_{b1}^{sl}+2\mathcal{G}_{b2}^{sl}}{\tilde{G}_a},\label{eq:new1}
\end{equation}
Equation (\ref{eq:new1}) shows that to lowest order in spin-flip scattering there are only two relevant parameters for the interface in a periodic N$_1$/N$_2$ multilayer with diffusive layers: its renormalized conductance $\tilde{G}$ and the symmetric spin-loss conductance $\mathcal{G}^{sl}=(\mathcal{G}_{1}^{sl}+\mathcal{G}_{2}^{sl})/2$. Under these conditions, the treatment based on the Valet-Fert model, with $\delta$ given by Eq.\ (8) of the main text, gives the same result as the full circuit theory. This justifies our treatment in the main text, where the correspondence with the Valet-Fert model was established for a multilayer with vanishing bulk resistance and spin relaxation.

Higher-order correction to ${\cal D}$ is always positive, which means that we have slightly overestimated $\delta$. However, this correction is very small for the Cu/Pd interface; for $\delta=0.4$ and typical parameters for bulk Pd \cite{S_Bass2015} the correction to $\delta^2$ is less than $0.01$. The correction may, however, be significant for interface with strong spin-flip scattering, such as Cu/Pt with $\delta\sim1$ \cite{S_Bass2015}.


\begin{thebibliography}{99}
\bibitem{note-Pd} The ramp-up region is needed only on the Pd side of the Cu/Pd interface, where strong spin-flip scattering otherwise occurs at the sharp boundary between Pd without SOC (lead) and Pd with SOC (scattering region).
\bibitem{note-filtering} As will be reported elsewhere, a more accurate procedure based on filtering by $\mathbf{k}$-points tends to yield $\delta$ values that are slightly larger, by 10-15\%.
\end{thebibliography}
\end{document}